\documentclass[12pt]{article}

\usepackage{amsmath, amsthm, amsfonts, amssymb}
\newtheorem{thm}{Theorem}[section]
\newtheorem{cor}[thm]{Corollary}
\newtheorem{lem}[thm]{Lemma}
\newtheorem{prop}[thm]{Proposition}
\newtheorem{rem}[thm]{Remarque}
\newtheorem{defn}[thm]{Definition}

\newcommand{\semikfk}{$\left\{P^{D,V}_t\right\}_{t\geq 0}$}
\newcommand{\semifk}{$\left\{P^V_t\right\}_{t\geq 0}$}

\newcommand{\dem}{{\bf Proof.\ }}
\newcommand{\fin}{\rule{0.5em}{0.5em}}
\newcommand{\semi}{$\left\{T(t)\right\}_{t\geq 0}\;$}
\newcommand{\semia}{$\left\{T^{*}(t)\right\}_{t\geq 0}\;$}

\title{{\bf $L^\infty$-uniqueness of Schr\"odinger operators restricted in an open domain}\thanks{This
work is partially supported by {\it Yangtze Research Programme},
Wuhan University, China, and the {\it Town Council} of Hunedoara,
Romania.}}
\author
{Ludovic Dan LEMLE\thanks{UFR Sciences et Technologies, Universit\'e
Blaise Pascal, 63177 Aubi\`ere, France and Engineering Faculty, "Politehnica"
University, 331128 Hunedoara, Romania\quad e-mail: {\tt lemle.dan@fih.upt.ro}}
}

\date{revised version 20 January 2008}

\def\E{\mathbb{E}}
\def\K{\cal{K}}
\def\N{\mathbb{N}}
\def\P{\mathbb{P}}

\def\R{\mathbb{R}}

\begin{document}

\maketitle

\begin{abstract}
\noindent Consider the Schr\"odinger operator ${\cal
A}=-\frac{\Delta}{2}+V$ acting on space $C_0^\infty(D)$, where
$D$ is an open domain in $\R^d$. The main purpose of this paper is
to present the $L^\infty(D,dx)$-uniqueness for Schr\"odinger
operators which is equivalent to the $L^1(D,dx)$-uniqueness of weak
solutions of the heat diffusion equation associated to the operator
$\cal A$.
\end{abstract}


\section{Preliminaries}

Let $D$ be an open domain in $\R^d$ with its boundary $\partial D$.
We denote by $C_0^\infty(D)$ the space of all infinitely
differentiable real functions on $D$ with compact support. Consider
the {\it Schr\"odinger operator} ${\cal A}=-\frac{\Delta}{2}+V$ acting on space $C_0^\infty(D)$, where $\Delta$ is the Laplace operator
and $V:\R^d\longrightarrow\R$ is a Borel measurable potential.\\ 
The {\it essential self-adjointness} of Schr\"odinger operator in
$L^2\left(\R^d,dx\right)$, equivalent to the unique solvability of
Schr\"odinger equation in $L^2\left(\R^d,dx\right)$, has been studied by
{\sc Kato} \cite{kato'84}, {\sc Reed} and {\sc Simon}
\cite{reed-simon'75}, {\sc Simon} \cite{simon'82} and others because
of its importance in Quantum Mechanics. In the case where $V$ is bounded, it is not difficult to prove
that $\left({\cal A},C_0^\infty(\R^d)\right)$ is essentially self-adjoint in 
$L^2(\R^d,dx)$. But in almost all interesting situations in
quantum physics, the potential $V$ is unbounded. In this situation we
need to consider the {\it Kato class}, used first by {\sc
Schechter} \cite{schechter'71} and {\sc Kato} \cite{kato'72}. A real valued measurable function $V$ is said to be in the {\it Kato
class} ${\K}^d$ on $\R^d$ if
$$
\lim_{\delta\searrow
0}\sup_{x\in\R^d}\int\limits_{|x-y|\leq\delta}\left|g(x-y)V(y)\right|dy=0
$$
where
$$
g(x)=\left\{\begin{array}{lcl}
\frac{1}{|x|^{d-2}}&,&\mbox{ if }d\geq 3\\
\ln\frac{1}{|x|}&,&\mbox{ if }d=2\\
1&,&\mbox{ if }d=1.
\end{array}
\right.
$$
If $V\in L^2_{loc}\left(\R^d,dx\right)$ is such that
$V^{-}$ belongs to the Kato class on $\R^d$, it is well known that the Schr\"odinger 
operator $({\cal A},C_0^\infty(\R^d))$ is essentially self-adjoint and the unique solution in $L^2$ of the heat equation is given by the famous {\it Feynmann-Kac semigroup} \semifk
$$
P^V_tf(x):=\E^xf(B_t)exp\left(-\int\limits_0^t\!V(B_s)\:ds\right)
$$ 
where $f$ is a nonnegative measurable function, $\left(B_t\right)_{t\geq 0}$ is the Brownian
Motion in $\R^d$ defined on some
filtered probability space $\left(\Omega,{\cal F},\left({\cal
F}_t\right)_{t\geq 0},\left(\P_x\right)_{x\in\R^d}\right)$ with
$\P_x\left(B_0=x\right)=1$ for any initial point  $x\in\R^d$ and $\E^x$ means the expectation with respect to $\P_x$.\\
In the case where $D$ is a
strict sub-domain, sharp results are known only when $d=1$ or, in
the multidimensional case, only in some special situations.\\
Consequently of an intuitive probabilistic interpretation of
uniqueness, {\sc Wu} \cite{wu'98} introduced and studied the
uniqueness of Schr\"odinger operators in $L^1\left(D,dx\right)$. On say that $\left({\cal A}, C^\infty_0\left(D\right)\right)$ is $L^1\left(D,dx\right)$-unique if $\cal A$ is closable and its closure is the generator of some $C_0$-semigroup on $L^1\left(D,dx\right)$. This
uniqueness notion was also studied in {\sc Arendt} \cite{arendt'86}, {\sc
Eberle} \cite{eberle'97}, {\sc Djellout} \cite{djellout'97}, {\sc
R\"ockner} \cite{rockner'98}, {\sc Wu} \cite{wu'98} and
\cite{wu'99} and others in the Banach spaces setting.

\section{$L^\infty(D,dx)$-uniqueness of Schr\"odinger operators}

Our purpose is to study the
$L^\infty\left(D,dx\right)$-uniqueness of the Schr\"odinger operator
$\left({\cal A}, C^\infty_0\left(D\right)\right)$ in the case where $D$ is a
strict sub-domain on $\R^d$. But how we can define the uniqueness in $L^\infty(D,dx)$? One can prove rather easely that {\it the killed Feynmann-Kac} semigroup \semikfk
$$
P^{D,V}_tf(x):=\E^x1_{[t<\tau_D]}f(B_t)exp\left(-\int\limits_0^t\!V(B_s)\:ds\right)
$$
where $\tau_D:=inf\{t>0\::\:B_t\notin D\}$ is the first exiting time of $D$, is a semigroup of bounded operators on $L^p(D,dx)$ for any $1\leq p\leq\infty$, which is strongly continuous for $1\leq p<\infty$, but never strongly continuous in $(L^\infty(D,dx),\|\:.\:\|_\infty)$. Moreover, a well known result of {\sc Lotz} \cite[Theorem 3.6, p. 57]{lotz'86} says that the generator of any strongly continuous semigroup on $(L^\infty(D,dx),\|\:.\:\|_\infty)$ must be bounded.\\
To obtain a correct definition of $L^\infty(D,dx)$-uniqueness, we should introduce a weaker topology of $L^\infty(D,dx)$ such that \semikfk becomes a strongly continuous semigroup with respect to this new topology. Remark that the natural topology for studying $C_0$-semigroups on
$L^\infty\left(D,dx\right)$ used first by {\sc Wu} and {\sc Zhang} \cite{wu-zhang'06} is {\it the topology of uniform
convergence on compact subsets of $L^1\left(D,dx\right)$}, denoted
by ${\cal C}\left(L^\infty,L^1\right)$. More precisely, if we denote
$$
\left\langle f,g\right\rangle:=\int\limits_D\!f(x)g(x)dx
$$
for all $f\in L^1(D,dx)$ and $g\in\L^\infty(D,dx)$, then for an arbitrary point $g_0\in
L^\infty(D,dx)$, a basis of neighborhoods with respect to ${\cal
C}\left(L^\infty,L^1\right)$ is given by
$$
N(g_0;K,\varepsilon):=\left\{g\in L^\infty(D,dx)\::\:\sup_{f\in
K}\left|\left\langle f,g\right\rangle-\left\langle f,g_0\right\rangle\right|<\varepsilon\right\}
$$
where $K$ runs over all compact subsets of $L^1(D,dx)$ and
$\varepsilon>0$.\\
Remark that $(L^\infty(D,dx),{\cal
C}\left(L^\infty,L^1\right))$ is a locally convex space and if \semi is a $C_0$-semigroup on
$L^1\left(D,dx\right)$ with generator $\cal L$, by \cite[Tneorem 1.4, p. 564]{wu-zhang'06} it follows that \semia is a $C_0$-semigroup
on $\left(L^\infty(D,dx),{\cal C}\left(L^\infty,L^1\right)\right)$ with
generator ${\cal L}^{*}$.\\
Now we can introduce the uniqueness notion in $L^\infty(D,dx)$. Let $\bf A$ be a linear operator on $L^\infty(D,dx)$ with domain $\cal D$ wich is assumed to be dense in $L^\infty(D,dx)$ with respect to the topology ${\cal C}\left(L^\infty,L^1\right)$.
\begin{defn}
The operator $\bf A$ is said to be a pre-generator on
$L^\infty(D,dx)$ if there
exists some $C_0$-semigroup on $\left(L^\infty(D,dx),{\cal
C}\left(L^\infty,L^1\right)\right)$ such that its generator $\cal L$
extends $\bf A$. We say that $\bf A$ is $L^\infty(D,dx)$-unique if $\bf A$ is closable and its
closure with respect to the topology ${\cal C}\left(L^\infty,L^1\right)$ is the generator of some
$C_0$-semigroup on $\left(L^\infty(D,dx),{\cal C}\left(L^\infty,L^1\right)\right)$.
\end{defn}
The main result of this paper is
\begin{thm}\label{1.2}
Let $V\in L^\infty_{loc}\left(D,dx\right)$ such that
$V^{-}\in{\K}^d$. Then the Schr\"odinger operator $\left({\cal A},
C^\infty_0\left(D\right)\right)$ is $(L^\infty(D,dx),{\cal C}\left(L^\infty,L^1\right))$-unique.
\end{thm}
\dem
First, we must remark that the existence assumption of pre-generator in \cite[Theorem 2.1, p. 570]{wu-zhang'06} is
satisfied. Indeed, if consider the killed Feynman-Kac semigroup
\semikfk on $L^\infty\left(D,dx\right)$ and for any $p\in[1,\infty]$ we define 
$$
\left\|P^{D,V}_t\right\|_p:=\sup_{\stackrel{f\geq
0}{\|f\|_p\leq 1}}\left\|P^{D,V}_tf\right\|_p,
$$
next lemma show that $\cal A$ is a pre-generator on $(L^\infty(D,dx),{\cal C}\left(L^\infty,L^1\right))$, i.e. $\cal A$ is contained in the generator ${\cal L}^{D,V}_{(\infty)}$ of the killed Feynmann-Kac semigroup \semikfk. 
\begin{lem}\label{3.2}
Let $V\in L^\infty_{loc}\left(D,dx\right)$ such that $V^{-}\in{\K}^d$ and let \semikfk be the
killed Feynman-Kac semigroup on $L^\infty\left(D,dx\right)$. If
$\left\|P^{D,V}_t\right\|_\infty$ is bounded over the compact
intervals, then \semikfk is a $C_0$-semigroup on
$\left(L^\infty(D,dx),{\cal C}\left(L^\infty,L^1\right)\right)$ and its
generator ${\cal L}^{D,V}_{(\infty)}$ is an extension of
$\left({\cal A},C^\infty_0(D)\right)$.
\end{lem}
\dem 
The proof is close to that of \cite[Lemma 2.3, p. 288]{wu'98}. Let \semikfk be the killed Feynman-Kac semigroup on $L^\infty(D,dx)$. 
Remark that
$$
\left|P^{D,V}_tf(x)\right|\leq P^{D,V}_t|f|(x)\leq
P^{D,-V^{-}}_t|f|(x)\leq P_t^{-V^{-}}|f|(x)
$$
from where we deduce that
$$
\sup_{0\leq t\leq 1}\left\|P^{D,V}_t\right\|_{\infty}\leq\sup_{0\leq
t\leq 1}\left\|P^{-V^{-}}_t \right\|_\infty<\infty
$$
since $\left\|P^{-V^{-}}_t \right\|_\infty$ is uniformly bounded by
the assumption that $V^{-}\in{\K}^d$ (see \cite{aizenman-simon'82}). Since $\left\|P^{D,V}_t\right\|_1=\left\|P^{D,V}_t\right\|_\infty$ is bounded for $t$ in compact intervals of $[0,\infty)$, using \cite[Lemma 2.3, p. 59]{wu'01} it follows that \semikfk is a $C_0$-semigroup on $L^1(D,dx)$. By \cite[Theorem 1.4, p. 564]{wu-zhang'06} we find that \semikfk is a $C_0$-semigroup on $L^\infty(D,dx)$ with respect to the topology ${\cal C}(L^\infty,L^1)$. We have only to show that its generator ${\cal L}^{D,V}_{(\infty)}$ is an extension of $\left({\cal A},C^\infty_0(D)\right)$.\\
{\bf Step 1: the case $V\geq 0$.} For $n\in\N$ we consider $V_n:=V\wedge n$. By a theorem of bounded perturbation (see \cite[Theorem 3.1, p. 68]{davies'80}) it follows that
$$
{\cal A}_n=-\frac{\Delta}{2}+V_n
$$
is the generator of a $C_0$-semigroup $\left\{P^{D,V_n}_t\right\}_{t\geq 0}$ on $\left(L^\infty(D,dx),{\cal C}\left(L^\infty,L^1\right)\right)$. So for any $f\in{\cal C}_0^\infty(D)$ we have
$$
P_t^{D,V_n}f-f=\int\limits_{0}^{t}P_s^{D,V_n}{\cal A}_nf\;ds\quad,\quad\forall t\geq 0.
$$
Letting $n\rightarrow\infty$, we have pointwisely on $D$:
$$
P_t^{D,V_n}f\rightarrow P_t^{D,V}f
$$
and
$$
P_t^{D,V_n}{\cal A}_nf\rightarrow P_t^{D,V}{\cal A}f\quad.
$$  
Moreover, for any $x\in D$ we have:
$$
\left|P_t^{D,V_n}f(x)\right|\leq P_t^{D,V}|f|(x)
$$
and
$$
\left|P_t^{D,V_n}{\cal A}_nf(x)\right|\leq P_t^{D,V}\left(\left|\frac{\Delta}{2}\right|+|Vf|\right)(x)\quad.
$$
Hence by the dominated convergence we derive that
$$
P_t^{D,V}f-f=\int\limits_{0}^{t}\!P_s^{D,V}{\cal A}fds\quad,\quad\forall t\geq 0.
$$
It follows that $f$ is in the domain of the generator ${\cal L}^{D,V}_{(\infty)}$ of $C_0$-semigroup \semikfk.\\ 
{\bf Step 2: the general case.} Setting $V^n=V\vee(-n)$, for $n\in\N$, and denoting by
$$
{\cal A}^n=-\frac{\Delta}{2}+V^n
$$
the generator of the $C_0$-semigroup $\left\{P^{D,V^n}_t\right\}_{t\geq 0}$ on $\left(L^\infty(D,dx),{\cal C}\left(L^\infty,L^1\right)\right)$, we have by Step 1
$$
P_t^{D,V^n}f-f=\int\limits_0^t\!P_s^{D,V^n}{\cal A}^nfds\quad,\quad t\geq 0.
$$
Notice that 
$$
\left|P_s^{D,V^n}{\cal A}^nf(x)\right|\leq P_s^{D,V}\left(\left|\frac{\Delta}{2}f\right|+|Vf|\right)(x)
$$
which is uniformly bounded in $L^\infty(D,dx)$ over $[0,t]$. By Fubini's theorem we have
$$
\int\limits_0^t\!P_s^{D,V}\left(\left|\frac{\Delta}{2}f\right|+|Vf|\right)(x)ds<\infty\mbox{  dx-a.e. on }D.
$$
On the other hand, for any $x\in D$ fixed such that
$$
P_s^{D,V}\left(\left|\frac{\Delta}{2}f\right|+|Vf|\right)(x)<\infty
$$
then by dominated convergence we find
$$
P_s^{D,V^n}\left(-\frac{\Delta}{2}+V^n\right)f(x)\longrightarrow P_s^{D,V}\left(-\frac{\Delta}{2}+V\right)f(x)\quad.
$$
Thus by dominated convergence we have dx-a.e. on $D$,
$$
\int\limits_0^t\!P_s^{D,V^n}\left(-\frac{\Delta}{2}+V^n\right)fds\rightarrow\int\limits_0^t\!
P_s^{D,V}\left(-\frac{\Delta}{2}+V\right)fds\quad,\quad\forall t\geq 0.
$$
The same argument shows that
$$
P_t^{D,V^n}f-f\rightarrow P_t^{D,V}f-f\quad.
$$
By consequence
$$
P_t^{D,V}f-f=\int\limits_0^t\!P_s^{D,V}\left(-\frac{\Delta}{2}+V\right)fds\quad,\quad\forall t\geq 0.
$$
Hence $f$ is in the domain of generator ${\cal L}^{D,V}_{(\infty)}$ of semigroup \semikfk. So ${\cal L}^{D,V}_{(\infty)}$ is an extension of the operator $\left({\cal A},C^\infty_0(D)\right)$ and the lemma is proved.\\
Next we prove the $L^\infty(D,dx)$-uniqueness of $\cal A$. By \cite[Theorem 2.1, p. 570]{wu-zhang'06}, we
deduce that the operator $\left({\cal A},
C^\infty_0\left(D\right)\right)$ is
$L^\infty\left(D,dx\right)$-unique if and only if for some
$\lambda$, the range $(\lambda I-{\cal
A})\left(C^\infty_0\left(D\right)\right)$ is dense in
$\left(L^\infty(D,dx),{\cal C}\left(L^\infty,L^1\right)\right)$. It is
enough to show that for any $h\in L^1\left(D,dx\right)$ which
satisfies the equality
$$
\left\langle h,(\lambda I+{\cal
A})f\right\rangle=0\quad,\quad\forall f\in C^\infty_0\left(D\right)
$$
it follows $h=0$.\\
Let $h\in L^1\left(D,dx\right)$ be such that for some
$\lambda$ one have
$$
\left\langle h,(\lambda I+{\cal
A})f\right\rangle=0\quad,\quad\forall f\in C^\infty_0\left(D\right)
$$
or
$$
(\lambda I+{\cal A})h=0\quad\mbox{ in the sense of distribution.}
$$
Since $V\in L^\infty_{loc}\left(D,dx\right)$, by applying
\cite[Theorem 1.5, p. 217]{aizenman-simon'82} we can see that $h$ is
a continuous function. By the mean value theorem due to {\sc
Aizenman} and {\sc Simon} \cite[Corollary 3.9, p.
231]{aizenman-simon'82}, there exists some constant $C>0$ such as
$$
|h(x)|\leq C\int\limits_{|x-y|\leq 1}\!|h(y)|\:dy\quad,\quad\forall
x\in D.
$$
As $V^{-}\in{\K}^d$, $C$ may be chosen independently of $x\in D$.
Since $h\in L^1(D,dx)$, it follows that $h$ is bounded and,
consequently, $h\in L^2(D,dx)$. Now by the $L^2(D,dx)$-uniqueness of
$\left({\cal A},C_0^\infty(D)\right)$ and \cite[Theorem 2.1, p. 570]{wu-zhang'06}, $h$ belongs to the domain of the generator ${\cal
L}^{D,V}_{(2)}$ of \semikfk on $L^2$ and
$$
{\cal L}^{D,V}_{(2)}h=\left(-\frac{\Delta}{2}+V\right)h=-\lambda
h\quad.
$$
Hence
$$
P^{D,V}_th=e^{-\lambda t}h\quad,\quad\forall t\geq 0.
$$
Let
$$
\lambda(D,V):=\inf_{f\in C_0^\infty(D)}\left\{\frac{1}{2}\int\limits_D\!|\nabla f|^2dx+Vf^2dx\::\:\|f\|_2\leq
1\right\}.
$$
be the lowest energy of the Schr\"odinger operator. If we take $\lambda<\lambda(D,V)$, then the last equality
is possible only for $h=0$, because
$\left\|P^{D,V}_t\right\|_2=e^{-\lambda(D,V)t}$ (see {\sc
Albeverio} and {\sc Ma} \cite[Theorem 4.1, p. 343]{albeverio-ma'91}).\fin
\begin{rem}
\em
Intuitively, to have $L^1\left(D,dx\right)$-uniqueness, the repulsive potential $V^{+}$ should grow rapidly to
infinity near $\partial D$, this means
$$
(C_1)\quad\quad\quad\P_x\left(\int\limits_0^{\tau_D}\!V^{+}(B_s)\:ds+\tau_D=\infty\right)=1\quad\mbox{for a.e. }x\in D
$$
so that a particle with starting point
inside $D$ can not
reach the boundary $\partial D$ (see \cite [Theorem 1.1, p. 279]{wu'98}).\\
By analogy with the uniqueness in $L^1(D,dx)$, the $L^\infty(D,dx)$-uniqueness of $\left({\cal
A},C_0^\infty(D)\right)$ means that a particle starting from the
boundary $\partial D$ can not enter in $D$. Unfortunately, here we have a problem: $L^\infty(D,dx)$-uniqueness of $\cal A$ is equivalent to the existence of a unique boundary condition for ${\cal A}^{*}$. It is well known that there are many boundary conditions (Dirichlet, Newmann, etc.). Remark that in the case of $L^1(D,dx)$-uniqueness of $\cal A$, the effect of the boundary condition for ${\cal A}^{*}$ is eliminated by the condition $(C_1)$ for potential. To find such condition in the case of $L^\infty(D,dx)$-uniqueness is very dificult. In this moment we can present here an interesting result from \cite{wu-zhang'06}:
\end{rem}
\begin{prop}
Let $D$ be a nonempty open domain of $\R^d$. If the Laplacian $(\Delta,C_0^\infty(D))$ is $L^\infty(D,dx)$-unique, then $D^{C}=\O$ or $D=\R^d$.
\end{prop}       
For the heat diffusion equation we can formulate the next result
\begin{cor}
If $V\in L^\infty_{loc}(\R^d,dx)$ and $V^{-}\in{\cal
K}^d$, then for every $h\in L^1(\R^d,dx)$, the heat diffusion equation
$$
\left\lbrace\begin{array}{l}
\partial_tu(t,x)=\left(-\frac{\Delta}{2}+V\right)u(t,x)\\
u(0,x)=h(x)
\end{array}
\right.
$$
has one $L^1(\R^d,dx)$-unique weak solution which is given by
$u(t,x)=P^{V}_th(x)$.
\end{cor}
\dem
The assertion follows by \cite[Theorem 2.1, p. 570]{wu-zhang'06} and Theorem \ref{1.2}. \fin
\vspace{1cm}

\noindent {\bf Acknowledgements.} I am grateful to
 Professor Liming WU for his kind invitation to Wuhan University, China,
 during May-June 2006 where this result was reported and for his valuable help and support. And I want to thank to anonymous reviewer for sugestions.

\bibliographystyle{plain}

\end{document}